\documentclass[aps,prd,showkeys,showpacs,amssymb,amsfonts,epsf,
preprintnumbers,nofootinbib,superscriptaddress,twocolumn]{revtex4}

\usepackage{color}
\usepackage[dvips]{graphicx}
\usepackage{bm,latexsym,amsmath,amssymb,amsfonts,color}

\newcommand\beq{\begin{eqnarray}}
\newcommand\eeq{\end{eqnarray}}
\newcommand*{\bA}{{\bar A}}
\newcommand*{\bF}{{\bar F}}
\newcommand*{\hq}{{\hat q}}

\begin{document}

\title{Domain wall universe in the Einstein-Born-Infeld theory}

\author{Bum-Hoon Lee}
\email{bhl"at"sogang.ac.kr}
\affiliation{
Department of Physics, Sogang University,
Shinsu-dong 1, Mapo-gu, Seoul 121-742, South Korea}
\affiliation{Center for Quantum Spacetime, Sogang University,
Shinsu-dong 1, Mapo-gu, Seoul 121-742, South Korea}

\author{Wonwoo Lee}
\email{warrior"at"sogang.ac.kr}
\affiliation{Center for Quantum Spacetime, Sogang University,
Shinsu-dong 1, Mapo-gu, Seoul, 121-742 South Korea}

\author{Masato Minamitsuji}
\email{minamituzi"at"sogang.ac.kr}
\affiliation{Center for Quantum Spacetime, Sogang University,
Shinsu-dong 1, Mapo-gu, Seoul, 121-742 South Korea}

\begin{abstract}
In this Letter, we discuss the dynamics of a domain wall universe
embedded into the charged black hole spacetime
of the Einstein-Born-Infeld (EBI) theory.
There are four kinds of possible spacetime structures,
i.e.,
those with no horizon,
the extremal one,
those with two horizons (as the Reissner-Nordstr$\rm{\ddot o}$m black hole),
and those with a single horizon (as the Schwarzshild black hole).
We derive
the effective cosmological equations on the wall.
In contrast to the previous works, we
take the contribution of the electrostatic energy on the wall into account.
By examining the properties of the effective potential,
we find that a bounce can always happen outside the
(outer) horizon. For larger masses of the black hole, the height of
the barrier between the horizon and bouncing point in
the effective potential becomes smaller, leading to longer time
scales of bouncing process.
These results are compared with those in the previous works.
\end{abstract}

\preprint{CQUeST-2009-0267}
\maketitle

\paragraph*{Introduction:}

The idea that our Universe corresponds to a topological defect
existing in a higher-dimensional spacetime has rather long history.
The pioneering phenomenological models have been proposed in
Refs. \cite{aka,rub}.
Recently,
motivated by the recent developments in string theory,
especially
by the discovery of objects (branes), where
the ordinary particles and interactions can be trapped,
these models have been extensively studied to explain various issues
in theoretical physics.
Phenomenologically interesting models were especially those
discussed in \cite{ADD,RS2}
(see e.g., \cite{Rubakov} for the review).
In the context of {\it brane world}, restrictions on the size of extra
dimensions become much weaker.
This idea has brought our
interests to the detection of the presence of the extra
dimensions, through collider experiments
or table-top tests on gravitational law.

The (2nd) model of Randall and Sundrum (RS) \cite{RS}, where a domain
wall (=brane) universe is placed at the orbifold fixed point of
the anti-de Sitter (AdS) space, has interesting properties for cosmology.
After the proposal of RS,
its application to cosmology has been studied by many authors.
In the thin domain wall approximation,
the dynamics of the domain wall can be traced by the
junction conditions.
The resulting
cosmology strongly depends on the contents of the matter
on the wall and the external geometry.
In the simplest AdS black hole spacetime,
with the $Z_2$ symmetry across the wall, the effective Friedmann
equations are the same as the conventional ones, except for two
important modifications \cite{bdl, kraus}. The first one is the
term proportional to the square of the energy density on the
wall and the other is a radiation-type contribution
caused by
a geometrical effect. These two effects could induce
modifications of cosmology in high energy regimes and have been
constrained from observations. It can be naturally expected that
in the more general spacetime, there would be new geometrical
effects, which may explain the origins of dark matter and dark
energy.
The solutions of inflationary domain walls
in the setup of the RS model
were obtained in Ref. \cite{kp,nh,hbkhdk}.

In this Letter,
we will reconsider the problem about the
dynamics of a domain wall universe in the charged
black hole spacetime.
This topics has been discussed not only in the context of
Einstein-Maxwell (EM) theory in Ref. \cite{rn1,rn2,rn3},
but also in that of
the Einstein-Born-Infeld (EBI) theory in \cite{bi1,bi2},
where electromagnetic fields with the BI Lagrangian
are coupled to gravity.
The BI theory is a
kind of generalizations of the Maxwell theory
\cite{bi_theory} and contains infinite number
of higher derivative terms of gauge potential,
which can be written in terms of the square root form.
The special property of
the BI theory
is that the electrostatic energy of a
point charge becomes finite.
Thus, it would be a good candidate for
the UV complete theory of the gauge field
(see e.g., \cite{zwb} for reviews).
The black hole solutions in the EBI theory have been studied
in \cite{cg,bibh,bibh2}.
Thermodynamical properties of
EBI black holes solutions
have been studied in Refs.~\cite{mo,yyy,mkp,crp}.
The vortices and
monopoles were studied,
in Ref. \cite{mns} and Ref. \cite{hk}, respectively.
In string theory,
the BI term describes the electromagnetic fields
living on the worldvolumes of $D$-branes.
Applications of AdS-BI black hole solutions to the
gauge-gravity duality have been discussed in \cite{tan}.


In the previous works, it has been pointed out that
a domain wall universe would experience a regular bounce in both of
the EM \cite{rn1,rn2,rn3} and the EBI models \cite{bibh,bi1,bi2}.
However, in these
works, the electrostatic energy stored on the wall charge has not been
taken into account. In the charged black hole spacetime, the
domain wall should be charged to generate the jump of the
electric field across the wall. The electrostatic energy of the
wall can contribute to its dynamics. $U(1)$-junction condition,
which can be derived via the variation of total action with
respect to the bulk gauge field, determines the amount of the
charge stored on the wall.

\vspace{0.2cm}

\paragraph*{Set-up:}
We consider the dynamics of an $n$-dimensional domain wall $\Sigma$
embedded into the $(n+1)$-dimensional spacetime $M_{\pm}$, where
the indices $(\pm)$ denote the left and right sides with respect
to the wall. The bulk action contains the matter as
well as the gravitation with a negative cosmological constant
while the wall action can contain the arbitrary kind of matter as
well as the tension.
Therefore, the total action is given by
\beq S&=&
\sum_{I=\pm} \int_{M_I} d^{n+1}X \sqrt{-G_I}
\Big[\frac{1}{2\kappa_{n+1}^2}
   \Big({}^{(n+1)}R_I-2\Lambda_I\Big)
  +  {\cal L}_I
\Big]
\nonumber \\
&+&\int_{\Sigma} d^nx
\sqrt{-g}
\Big(-\sigma+{\cal L}_m\Big)
\nonumber\\
&+&\frac{1}{\kappa_{n+1}^2}\int_{\Sigma}
 d^{n}x \sqrt{-g}
\Big( K_{+}+K_-\Big),
\label{theory}
\eeq
where
${\cal L}_{I}$ and ${\cal L}_m$ are matter and fields
living in each bulk region and on the wall, respectively.
The final term represents the Gibbons-Hawking term \cite{gh}.

We assume ${\cal L}_I$ as the electromagnetic field
with the BI Lagrangian.
Noth that
only the electric components of
the gauge field will appear in our problem.
Thus, the bulk matter Lagrangian is chosen to be
\beq
 {\cal L}_{I}
=
4
\alpha^2
\left(
1-\sqrt{1+\frac{F^{(I)}_{MN}F^{MN}_{(I)}}{8
\alpha^2}}
\right), \eeq where $F_{MN}=\partial_M A_N-\partial_N A_M$ is the
$U(1)$ field strength and $A_M$ is the corresponding vector
potential. The BI theory is a kind of generalization of Maxwell theory
and the parameter $\alpha$ controls
the degree of deviation from
the Maxwell theory.
It
is known that in the BI theory the electrostatic energy of a point
particle is finite and is a good for successful UV completion to
the Maxwell theory. In $\alpha\to \infty$ limit, the Maxwell
theory is recovered $ {\cal L}_{\pm}\approx
-F^{(\pm)}_{MN}F^{MN}_{(\pm)}/4$.
The electric flux density is given by the nonvanishing components
of the tensor defined by
\beq E_{AB}=\frac{F_{AB}}
{\sqrt{1+\frac{F_{MN}F^{MN}}{8\alpha^2}}}.
\eeq
By defining this,
one can generalize Gauss's law in the Maxwell theory,
to the case of the BI theory.
For our purpose, it
is rather convenient to introduced
the
rescaled parameter and
gauge field
by
$\beta:=\sqrt{2}\kappa_{n+1}\alpha$ and ${\bar
A}_{M}=\sqrt{2}\kappa_{n+1} A_{M}$,
respectively.
\vspace{0.2cm}


\paragraph*{Domain wall universe:}
Then, we discuss the dynamics of a domain wall universe. It is
assumed that the wall is infinitely thin and the Israel junction
conditions can be applied. The bulk metric has a maximally
symmetric, external $(n-1)$-dimensional space and one static extra
dimension:
\beq
ds_{\pm}^2=-f_{\pm}(R)
dT_{\pm}^2+R^2 \gamma_{ij}dx^i dx^j
+\frac{dR^2}{f_{\pm}(R)}\,,\label{yogi}
\eeq
where $\gamma_{ij}$
is the metric of a maximally symmetric $(n-1)$-dimensional space
and the subscripts $(\pm)$ specify the bulk region.

One may choose the common spatial coordinates $R=R_+=R_-$
and $x^i_+=x^i_-=x^i$.
The domain wall is located at $(R=a(\tau),T_{\pm}=T_{\pm}(\tau))$, where
$\tau$ represents the proper time on the wall,
defined by
$-f_{\pm}(a) \dot{T}_{\pm}^2+\dot{a}^2/f_{\pm}(a)=-1$.
$\epsilon_{\pm}=+ 1(-1)$ represents the outward (inward) direction,
i.e., that of increasing (decreasing) $a$
with respect to the wall.
The induced metric on the wall becomes
the Friedmann-Robertson-Walker form of curvature $K$
with the scale factor $a$.

By variating the action (\ref{theory}) with respect to the
metric degrees of freedom, the dynamics of a wall is determined by
the Israel junction condition \cite{israel}:
$\Big[K_{\mu\nu} -g_{\mu\nu}K \Big] =-\kappa_{n+1}^2 S_{\mu\nu}$,
where the stress-energy tensor on the wall is defined by
\beq
\sqrt{-g}S^{\mu\nu} =2\frac{\delta }{\delta g_{\mu\nu}}
 \int d^n x\sqrt{-g}
  \Big(-\sigma+{\cal L}_m\Big),
\eeq
and $[A]:= A_+-A_-$ represents the jump of a bulk quantity
$A$ across the wall. In general, the wall induced stress-energy
tensor in general has the form $ S^{\mu}{}_{\nu}={\rm diag}
\big(-\rho-\sigma,p-\sigma,\cdots,p-\sigma\big) $. $\sigma$,
$\rho$ and $p$ are the wall tension, the energy density and
pressure of the time-dependent matter, respectively. The nonvanishing
components of Israel junction conditions are given by
\beq &&-\frac{n-1}{a}\sum_{I}
 \epsilon_I\sqrt{f_I+\dot{a}^2}
=\kappa_{n+1}^2\Big(\rho+\sigma \Big),
\nonumber\\
&& \sum_{I} \epsilon_I \Big( \frac{n-2}{a}\sqrt{f_I +\dot{a}^2}
+\frac{f_{I,a}}{2}
 \frac{1}
      {\sqrt{f_I+\dot{a}^2}}
\nonumber\\
&+&
\frac{\ddot{a}}{\sqrt{f_I +\dot{a}^2}} \Big)
=\kappa_{n+1}^2\Big(p-\sigma\Big)\,,\label{konishi}
\eeq
where
the index $I$ runs $(\pm)$ and we defined
$\bF_{(I)}^2:=\bF^{(I)}_{MN}\bF^{MN}_{(I)}$.

Then, we derive the $U(1)$ junction condition
across the wall. The variation of the total action
(\ref{theory}) with respect to $A_M$ gives
\beq \delta_{\bA} S
&=&\frac{1}{2\kappa_{n+1}^2} \sum_I \int_{M_I} d^{n+1}X\sqrt{-G_{(I)}}
\nonumber\\
&\times&
 \nabla_C\Big(
\frac{
 G_{(I)}^{AB}G_{(I)}^{EC}
    \bF^{(I)}_{AE}}{\sqrt{1+\bF_{(I)}^2/(8\beta^2})}\Big)\delta \bA^{(I)}_{B}
\nonumber\\
&-&\frac{1}{2\kappa_{n+1}^2}
\int_{\Sigma} d^{n}y\sqrt{-g}
\sum_I   \frac{
      n^{(I)}_C G_{(I)}^{AB}G_{(I)}^{EC}\bF^{(I)}_{EA}\delta \bA^{(I)}_B}
          {\sqrt{1+\bF_{(I)}^2/(8\beta^2)}}
\nonumber \\
&+&
\int_{\Sigma} d^{n}y\sqrt{-g}\frac{\delta {\cal L}_m}{\delta \bA_{\tau}}
\delta \bA_{\tau}.
\eeq
Here on the boundary, we must impose the continuity condition
of the gauge potential:
$\bA^{(+)}_{M}u_{(+)}^M=\bA^{(-)}_{M}u_{(-)}^M=:\bA_{\tau}$ on
the wall $\Sigma$.
The bulk parts of the variation $\delta \bA_{(\pm)}$ in
$M_{(\pm)}$ give rise to the equation of motion
\beq
\nabla_C \left(
 \frac{    \bF^{(I)BC}}
{\sqrt{1+\bF_{(I)}^2/(8\beta^2)}} \right) =0.
\eeq
Then, we
derive the condition that the bulk gauge field satisfy on the
boundary.
In our spacetime,
on $\Sigma$,
$n_R=\epsilon \dot{T}$, $u^T=\dot{T}$
and $n^{(\pm)}_E G_{(I)}^{AB}G_{(I)}^{CE}
 \bF^{(I)}_{EA}\delta \bA^{(I)}_B
=\epsilon_{I} \bF^{(I)}_{T_{I}R} \delta \bA_{\tau}$.
Thus $U(1)$-junction condition on $\Sigma$ is given by
\beq
\sum_I   \frac{
\epsilon_{I} \bF^{I}_{T_I R}}
        {\sqrt{1+\frac{\bF_{I}^2}{8\beta^2}}}
=2\kappa_{n+1}^2
\frac{\delta {\cal L}_m}{\delta \bA_{\tau}}.\label{mac}
\eeq

\vspace{0.2cm}


\paragraph*{Born-Infeld black hole:}
We briefly review the static black hole solutions
in the EBI theory.
We will forces on the case of the asymptotically AdS
bulk spacetime $\Lambda<0$.
For the moment, we omit the subscripts $(\pm)$.
The metric is given by
\beq
ds^2
=-f(R) dT^2
+\frac{dR^2}{f(R)}+R^2\gamma_{ij}dx^i dx^j,
\eeq
where
\beq
f(R)
&:=&
K-\frac{m^2}{R^{n-2}}
+\Big(\frac{4\beta^2}{n(n-1)}+\frac{1}{\ell^2}\Big)R^2
\nonumber\\
&-&\frac{2\sqrt{2}\beta}{n(n-1)R^{n-3}}
\sqrt{2\beta^2 R^{2n-2}
   + (n-1)(n-2)q^2}
\nonumber\\
&+&\frac{2(n-1)q^2}{nR^{2n-4}}
\nonumber\\
&\times&
{}_2F_1
\Big[
\frac{n-2}{2n-2},
\frac{1}{2},
\frac{3n-4}{2n-2};
-\frac{(n-1)(n-2)q^2}{2\beta^2 R^{2n-2}}
\Big].
\label{ryuk}
\eeq
and
${}_2F{}_1[a,b,c;x]$ is Gauss's hypergeometric function.
The AdS curvature length scale is related to the bulk cosmological constant
through
$\ell:=\sqrt{-(n-1)(n-2)/\Lambda}$.

In the Maxwell limit $\beta\to \infty$
and/or for larger $R$,
the solution reduces to the AdS Reissner-Nordstr$\rm{\ddot o}$m
(AdS-RN) one:
\beq
f(R)
=
K+\frac{R^2}{\ell^2}+\frac{q^2}{R^{2n-4}}
-\frac{m^2}{R^{n-2}}
+O\Big(\frac{1}{R^{4n-6}}\Big).
\eeq
In the regime $\beta = O(1)$, however,
the behavior of the metric function can be modified
because
\beq
f(R)
&=&
K-\frac{m^2-A(n,\beta, q)}{R^{n-2}}
\nonumber\\
&-&\Big[\frac{2c(n)\beta}{n}-B(n,\beta,q)(2n-1)q\Big]\frac{q}{R^{n-3}}
\nonumber\\
&+&\Big[\frac{4\beta^2}{n(n-1)}+\frac{1}{\ell^2}\Big]R^2
+O(R^{n+1}),
\eeq
where
\beq
A(n,\beta,q)&:=&\frac{2(n-1)q^2}{n\sqrt{\pi}}
  \Big(\frac{2\beta^2}{(n-1)(n-2)q^2}\Big)^{(n-2)/(2n-2)}
\nonumber\\
&\times&
  \Gamma\Big(\frac{3n-4}{2n-2}\Big)
  \Gamma\Big(\frac{1}{2n-2}\Big),\label{inf}
\nonumber\\
c(n)&:=&\sqrt{\frac{2(n-2)}{n-1}}
\nonumber\\
B(n,\beta,q)&:=&\frac{4\beta}{cn(2n-1)q}
\frac{\Gamma\big(\frac{3n-4}{2n-2}\big)\Gamma\big(\frac{-1}{2n-2}\big)}
     {\Gamma\big(\frac{n-2}{2n-2}\big)\Gamma\big(\frac{2n-3}{2n-2}\big)},
\eeq
which are simply function of the charge $q$ other than the parameters
of the theory.
The nonvanishing component of $U(1)$ field strength is given by
\beq
\bF^{RT}
=\frac{2\sqrt{(n-1)(n-2)}\beta q}
             {\sqrt{2\beta^2 R^{2n-2}+(n-1)(n-2)q^2}}.
\label{youko}
\eeq
The black hole is now positively charged.
The corresponding gauge potential is given by
\beq
\bA_T&=&\Phi+
\sqrt{\frac{n-1}{2(n-2)}}
\frac{q}{R^{n-2}}
\nonumber\\
&\times&
{}_2F{}_1
\Big[
\frac{2(n-2)}{(n-1)},\frac{1}{2},\frac{3n-4}{2(n-1)},
-\frac{(n-1)(n-2)q^2}{2\beta^2 R^{2n-2}}
\Big]
\label{ola1}
\eeq
where $\Phi$ represents the possible constant shifts
of the vector potential.
In the vicinity of the black hole, $\bA_T$ is regular in contrast
to the Maxwell theory, while
in the far region, it falls off as $1/R^{n-2}$.
The modified Gauss's law determines the black hole charge
\beq
Q=\int \Big(4{\cal L}'(F^2) \Big)
\bF^{RT} R^{n-1} d\Omega_{(n-1)}.
\eeq
By performing the integration,
the total charge is found to be
$Q=\sqrt{2(n-1)(n-2)} q\Omega_{n-1}$
where $\Omega_{n-1}$ is the volume of $(n-1)$-sphere.
For the cases that $K=0$ and $K=-1$
Gauss's law can be satisfied per unit volume.

The position of the horizon can be written
as a function of other parameters
$R_{H}=G(\ell, q, m^2)$,
which can be inversely solved as
$m^2=F(\ell, q, R_H)$, where
\beq
F(\ell,q,R_H)
&=&K R_H^{n-2}
+\Big(\frac{4\beta^2}{n(n-1)}+\frac{1}{\ell^2}\Big)R_H^n
\nonumber\\
&-&\frac{2\sqrt{2}\beta R_H}{n(n-1)}
\sqrt{2\beta^2 R^{2n-2}
   + (n-1)(n-2)q^2}
\nonumber\\
&+&\frac{2(n-1)q^2}{nR_H^{n-2}}
\nonumber\\
&\times&
{}_2F_1
\Big[
\frac{n-2}{2n-2},
\frac{1}{2},
\frac{3n-4}{2n-2},
-\frac{(n-1)(n-2)q^2}{2\beta^2 R_H^{2n-2}}
\Big].
\nonumber\\
\eeq
If the equation $m^2=F(\ell,q,R_H)$ has
two, one and no roots of $R_H$
for a given mass, then the black hole has two, one and no horizons,
respectively.
In the extremal case,
two horizons are degenerating
at $R_H^{\rm ext}$,
where $\partial F/\partial R_H=0$ is satisfied.
In the case that $K=0$, the solution for the extremal condition
is given by
\beq
R_H^{\rm ext}
=\left(
 \frac{\beta\ell^2q\sqrt{8(n-2)}}
      {\sqrt{n}\sqrt{8\beta^2\ell^2+n(n-1)}}
 \right)^{1/(n-1)}\,,\label{ext}
\eeq
whose corresponding mass is given by
\beq
&&m_{\rm ext}^2
=\frac{2(n-1)q^2}{n}
\left(
\frac{\sqrt{n(8\beta^2\ell^2+n^2-n)}}
{\beta \ell^2\sqrt{8(n-2)}q}
\right)^{\frac{n-2}{n-1}}
\nonumber  \\
&\times&
{}_2F_1
\Big[
\frac{n-2}{2n-2},
\frac{1}{2},
\frac{3n-4}{2n-2},
-\frac{n(n-1)\big(8\beta^2\ell^2+n(n-1)\big)}
      {16\beta^4\ell^4}
\Big].
\eeq
In the limit $\beta\to \infty $,
\beq
m_{\rm ext}^2
\to \frac{2(n-1)q^{n/(n-1)}}{n\ell^{(n-2)/(n-1)}}
\Big(\frac{n}{n-2}\Big)^{\frac{n-2}{2(n-1)}},
\eeq
which is the result in the case of the
RN solution.
In the cases of curvature $K=\pm 1$,
there is no analytic way to express
$m_{\rm ext}^2$.

There would be at most two event horizons.
Spacetime has two event horizons if its mass is in the range
given by $m^2_{\rm ext}<m^2< A(q)$,
where $A$ is defined in Eq. (\ref{inf}).
It is straightforward to check that the ratio $m_{\rm ext}^2/A$
is always less than unity.
Similarly, the spacetime has a single horizon
for the larger masses $m^2>A(q)$
and
no horizon is formed
for the smaller (even negative)
masses $m^2<m_{\rm ext}^2$, i.e.,
a naked singularity at the center appears.
In the case that $K<0$,
for some sets of parameters
the mass of the black hole in the extremal case
can be negative
$m^2_{\rm ext}<0$.
In such a case, two horizons appear even
for negative mass of black holes.
In the limit that $\beta\to \infty$ we obtain $A\to \infty$
and there are always two horizons.
If the extremal condition (\ref{ext})
is not satisfied at any mass of the black hole,
only a single horizon appears.
In the case that $m^2<A$, no horizon is formed and
a naked singularity appears.
In the case that $m^2>A$, a single horizon is formed.


\vspace{0.2cm}

\paragraph*{Wall charge:}
The metric function $f_{\pm}$ in Eq. (\ref{yogi})
is given by Eq. (\ref{ryuk}), by replacing
$\ell$, $q$ and $m$
with
$\ell_{\pm}$, $q_{\pm}$ and $m_{\pm}$, respectively.
Now each bulk region is bounded by $0<R<a(\tau)$ for $\epsilon=-1$
and $R>a(\tau)$ for $\epsilon=+1$.
In the case $\epsilon=+1$, the graviton would not be localized
on the wall,
because the volume of the bulk is infinite
and graviton zero mode is not nonmailable.
The nonvanishing component of $U(1)$-gauge field
in each bulk side is
obtained by replacing
$\ell$,
$q$
and $m$
with $\ell_{\pm}$, $q_{\pm}$ and $m_{\pm}$
in Eq. (\ref{youko}).
The corresponding components of gauge potential is obtained
from Eq. (\ref{ola1}) by the same replacements
with $\Phi\to \Phi_{\pm}$.
$\Phi_{\pm}$ should be chosen to satisfy the continuity of the
wall component of the gauge potential:
$\bA_{\tau}=\bA^{(+)}_{T_+}\dot{T}_+
= \bA^{(-)}_{T_-}\dot{T}_-$.
The brane matter action is composed of
the part of the ordinal matter,
which is not coupled to the bulk electric field,
and
the one coupled to it:
\beq
 {\cal L}_m={\cal L}_0+
\frac{C}{a^{n-1}}\bA_{\tau},
\label{bic}
\eeq
where ${\cal L}_0$ represents the ordinary matter.
Then,
from the $U(1)$-junction condition across the wall (\ref{mac}),
we find
\beq
(2\kappa_{n+1}^2)C=\sqrt{2(n-1)(n-2)}
  \Big(\epsilon_+ q_+ +\epsilon_- q_-\Big).
\eeq In the $Z_2$ symmetric case $q_+=q_-=q$ and
$\epsilon_+=\epsilon_-=\epsilon$, \beq
(2\kappa_{n+1}^2)C=2\sqrt{2(n-1)(n-2)}\epsilon q. \eeq When
$\epsilon_{\pm}=-1$, the wall is negatively charged to neutralize
the positive charge of the black hole. In the case
$\epsilon_{\pm}=+1$, both sides of the bulk do not contain the
black hole horizons and electric field lines are extended from the
wall toward the infinity. In the case
$\epsilon_+=-\epsilon_-=+1$, the spacetime is infinitely extended
only into the $(+)$-direction. Then, the black hole charge is
neutralized by that of the
wall on the $(-)$ side, and the charge on the $(+)$-side generates
the electric field in the $(+)$-bulk. The opposite things happen
in the case that $\epsilon_+=-\epsilon_-=-1$.


\vspace{0.2cm}

\paragraph*{Cosmology:}
We assume that the bulk spacetime is $Z_2$-symmetric with respect to the wall:
$\ell:= \ell_+=\ell_-$, $q:= q_+=q_-$ and $m:=m_+=m_-$.
The energy density of the wall matter is given by
\beq
\rho
=\frac{2C\bA_T}{a^{n-1}f} \Big(f+\dot{a}^2\Big)^{1/2}+\rho_0
\eeq
where the first term represents the electrostatic energy
induced by the wall charge and
$\rho_0$ is that of the ordinary matter.
Note that the continuity condition
$\bA_{\tau}=\bA_T \dot{T}
=\bA_T\Big(f+\dot{a}^2\Big)^{1/2}/f$
and we may choose $\Phi=0$.
By squaring the first equation in Eq. (\ref{konishi}),
the effective cosmological equation can be derived
in the form in analogy with the classical mechanics
$\dot{a}^2+V(a)=0$,
where  we define the effective potential
\beq
V(a)&:=&
f(a)-
\frac{\kappa_{n+1}^4}{4(n-1)^2}
\frac{a^2 \big(\rho_0+\sigma\big)^2}
     {\big(1+G(a)\big)^2}.\label{pote}
\eeq
The function $G(a)$ is given by
\beq
G(a)=\sqrt{\frac{2(n-2)}{n-1}}
      \frac{ q \bA_T}{a^{n-2}f},
\eeq
which arises from the electrostatic energy.
For $n=4$, $G(a)\propto 1/a^7$ and
at the later time the domain wall cosmology approaches
the one in the neutral black hole background, including
the RS cosmology.

For the numerical visualization of the potential,
it is useful to
introduce the dimensionless quantities
$q=\hq \ell^{n-2}$,
$m=\hat m \ell^{(n-1)/2}$,
$\beta=\hat \beta\ell^{-1}$,
$t=\hat t \ell$ and
$a=\hat a \ell$.
The cosmological equations is reduced to
$\big(\hat a_{,\hat t}\big)^2
+\hat V(\hat a)
=0$,
where the potential is defined by
\beq
\hat V(\hat a)
= \hat f(\hat a)
 -\big(x\hat a\big)^2
   \frac{\Big(1+\frac{\rho_0}{\sigma}\Big)^2}
         {\big(1+\hat G(\hat a)\big)^2}.
\eeq
The dimensionless constant $x$ is introduced by
$x^2:=\kappa_{n+1}^2\big(\sigma \ell\big)^2/4/(n-1)^2$.
The case of the RS tuning is that $x=1$
and
we will focus on $x\geq 1$,
hence the universe approaches de Sitter geometry for larger $\hat a$.
The dimensionless function is defined by
\beq
\hat G(\hat a):=\sqrt{\frac{2(n-2)}{(n-1)}}
     \frac{{\hat q }{\hat A}_T({\hat a})}
          {\big({\hat a}\big)^{n-2}\hat f({\hat a})}
\eeq
where
\beq
{\hat f}({\hat a})
&=&K-\frac{\hat m^2}{\big(\hat a\big)^{n-2}}
+\Big(\frac{4\hat \beta^2}{n(n-1)}+1\Big)
\big(\hat a\big)^2
\nonumber\\
&-&\frac{2\sqrt{2}\hat \beta}{n(n-1)}
\sqrt{2\big(\hat \beta\big)^2\big(\hat a\big)^{2n-2}
+(n-1)(n-2)\hat q^2}
\nonumber\\
&+&
\frac{2(n-1)\big(\hat q\big)^2}
     {n \hat a^{2n-4}}
\nonumber\\
&\times&
{}_2F{}_1
\Big[
\frac{n-2}{2(n-1)},
\frac{1}{2},
\frac{3n-4}{2(n-1)},
-\frac{(n-1)(n-2)\hat q^2}
      {2\hat \beta^2\big(\hat a\big)^{2n-2}}
\Big],
\nonumber\\
&&
\eeq
and
\beq
\hat A_T&=&
\sqrt{\frac{n-1}{2(n-2)}}
\frac{\hat q}{\hat a^{n-2}}
\nonumber\\
&\times&
{}_2F{}_1
\Big[
\frac{2(n-2)}{(n-1)},\frac{1}{2},\frac{3n-4}{2(n-1)},
-\frac{(n-1)(n-2){\hat q}^2}{2{\hat \beta}^2 {\hat a}^{2n-2}}
\Big].
\label{ola3}
\eeq

In Fig. 1,
the behavior of $\hat f$ is shown as the function of $\hat a$
in the case of $K=0$
and for $n=4$, $\beta=10$ and $\hat q=3.0$.
(We also set $\rho_0=0$ for simplicity)
We consider the
case of no horizon ($m^2<m_{\rm ext}^2$),
the extremal case ($m^2=m_{\rm ext}^2$),
that of two horizons ($A>m^2>m_{\rm ext}^2$)
and that of a single horizon ($A<m^2$).
In Fig. 1, each case is described by
the solid, thick, dashed and dotted curves, respectively.
\begin{figure}
\begin{minipage}[t]{.45\textwidth}
   \begin{center}
    \includegraphics[scale=.75]{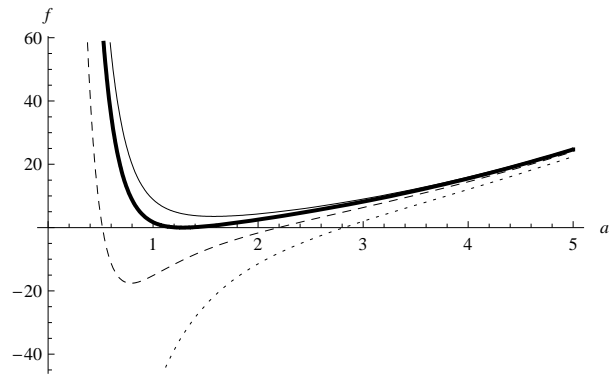}
\vspace{0cm}
        \caption{
Typical behaviors of the metric function $\hat f$ are shown as functions
of $\hat a$. The solid, thick, dashed and dotted curves correspond
to the cases of
no-horizon ($m^2<m_{\rm ext}^2$),
the extremal ($m^2=m_{\rm ext}^2$),
two-horizon ($A>m^2>m_{\rm ext}^2$),
single-horizon ($A<m^2$),
respectively.
We have chosen $n=4$, $\hat \beta=10$ and $\hat q=3$.
}
   \end{center}
 \end{minipage}
\end{figure}

In Fig. 2, for the same set of parameters and $x=1.1$,
the behavior of the
potential is shown as a function of $\hat a$.
From the top to the bottom,
the panels correspond to
the cases of no horizon,
extremal,
two-horizon
and single-horizon,
respectively.
\begin{figure}
\label{fig1}
\begin{minipage}[t]{.45\textwidth}
   \begin{center}
    \includegraphics[scale=.75]{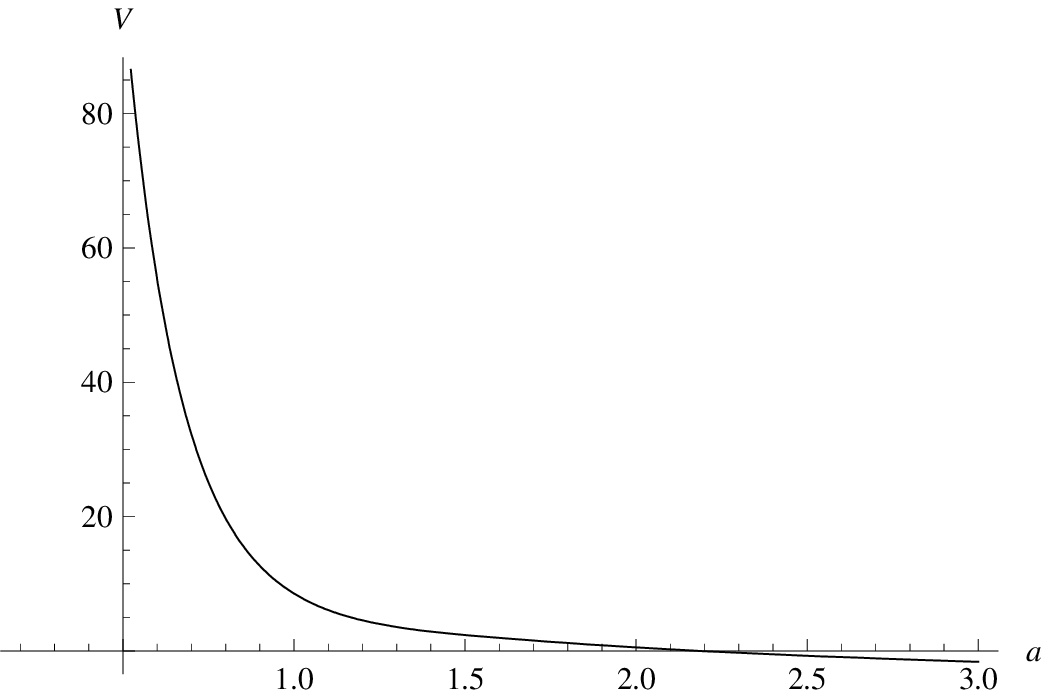}
    \includegraphics[scale=.75]{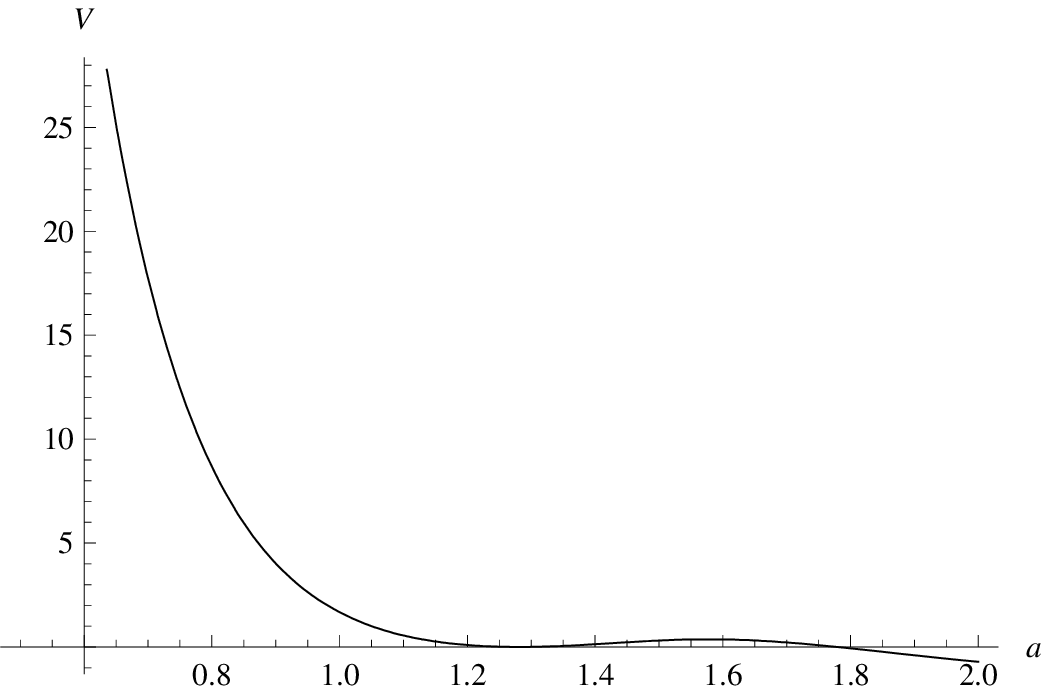}
    \includegraphics[scale=.75]{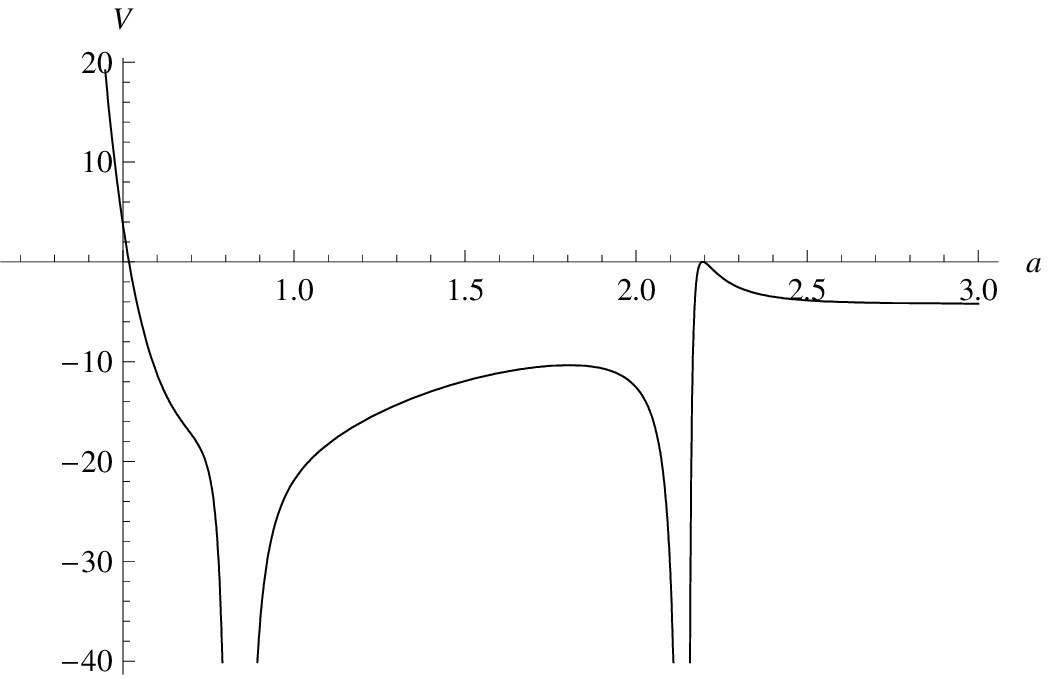}
    \includegraphics[scale=.75]{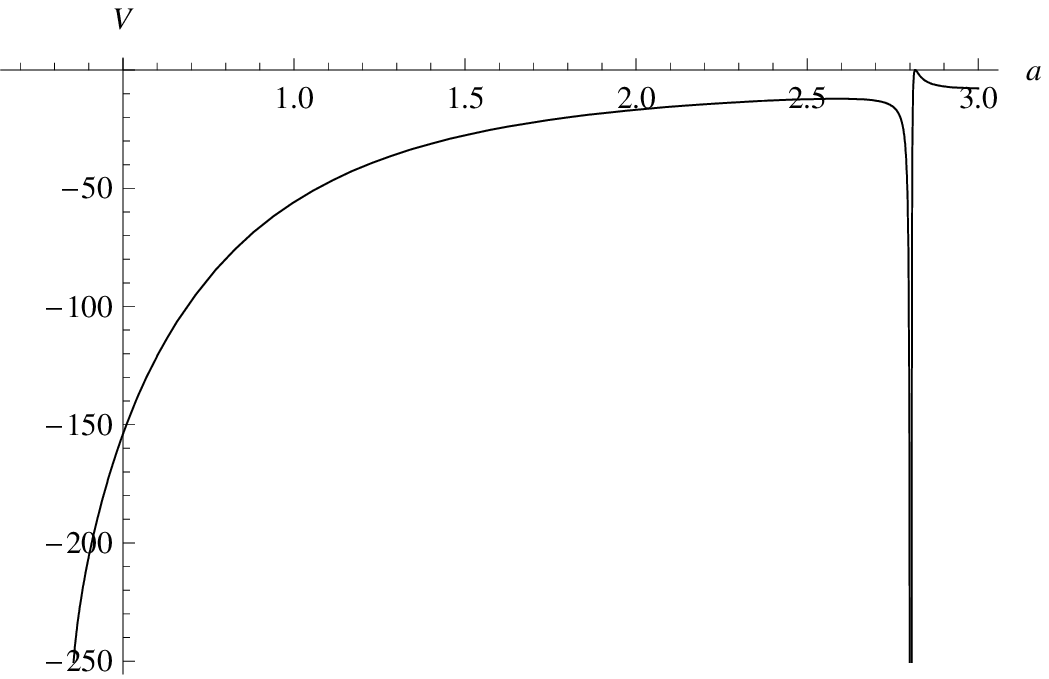}
\vspace{0cm}
        \caption{
Typical behaviors of the potential $\hat V$ are shown
as the function of $\hat a$.
The first, second, third and 4th panels show the cases of
no horizon ($m^2<m_{\rm ext}^2$),
the extremal ($m^2=m_{\rm ext}^2$),
two horizons ($A>m^2>m_{\rm ext}^2$)
and single horizon ($A<m^2$), respectively.
We have chosen $n=4$, $\hat \beta=10$, $\hat q=3$
and $x=1.1$.
}
   \end{center}
 \end{minipage}
\end{figure}
In the first two cases, the potential is regular everywhere
and the domain wall universe can undergo a bounce, if it
is outside the horizon.
In the last two cases, the potential negatively
diverges at some places inside the horizon.
But this divergence does not imply
the presence of any singularity of the
motion of the wall if it is inside the outer horizon, because
the velocity of the brane diverges just at this instance
and then becomes finite again.
In these cases, in order to
investigate its behavior just outside the horizon,
we have to enlarge our plots.

In Fig. 3,
for the same set of parameters as in the Fig. 1 and 2,
the near-horizon behaviors of the metric function and potential
are shown.
\begin{figure}
\begin{minipage}[t]{.45\textwidth}
   \begin{center}
    \includegraphics[scale=.75]{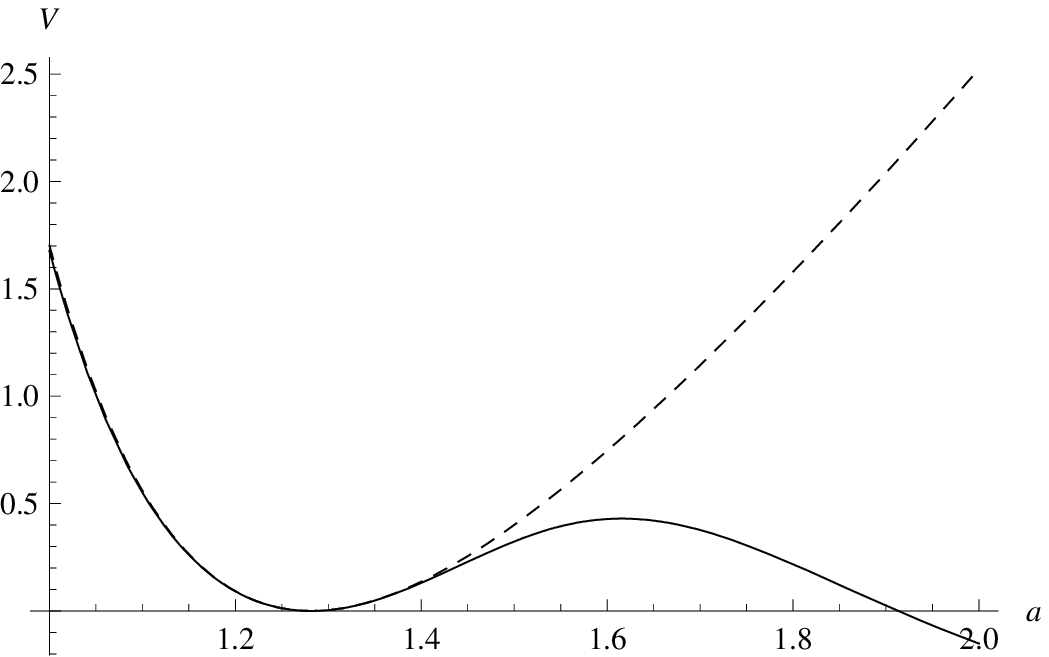}
    \includegraphics[scale=.75]{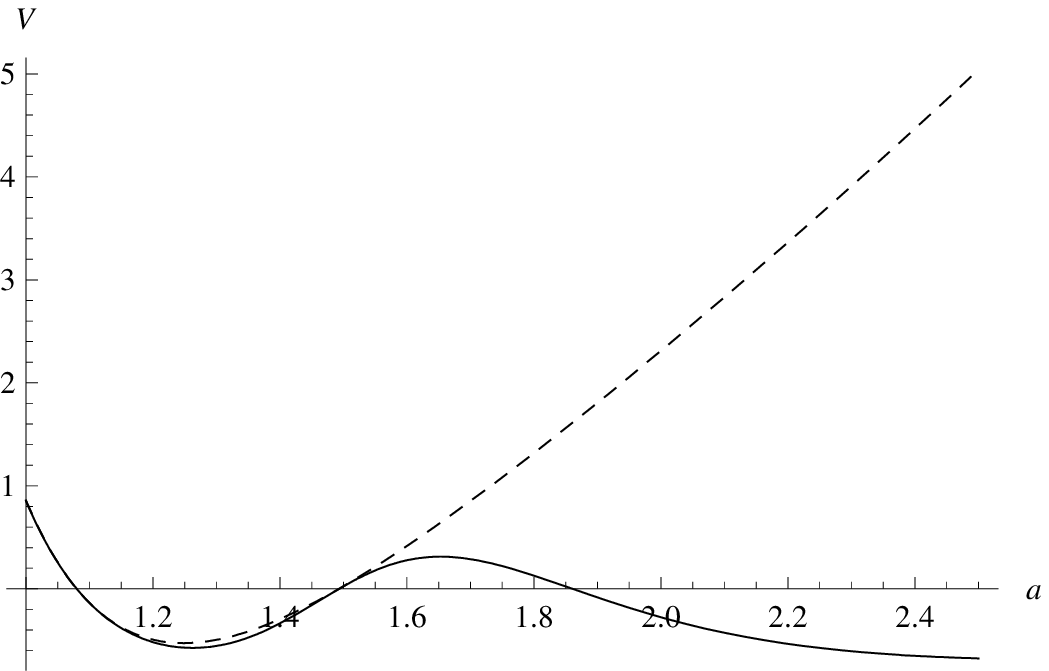}
    \includegraphics[scale=.75]{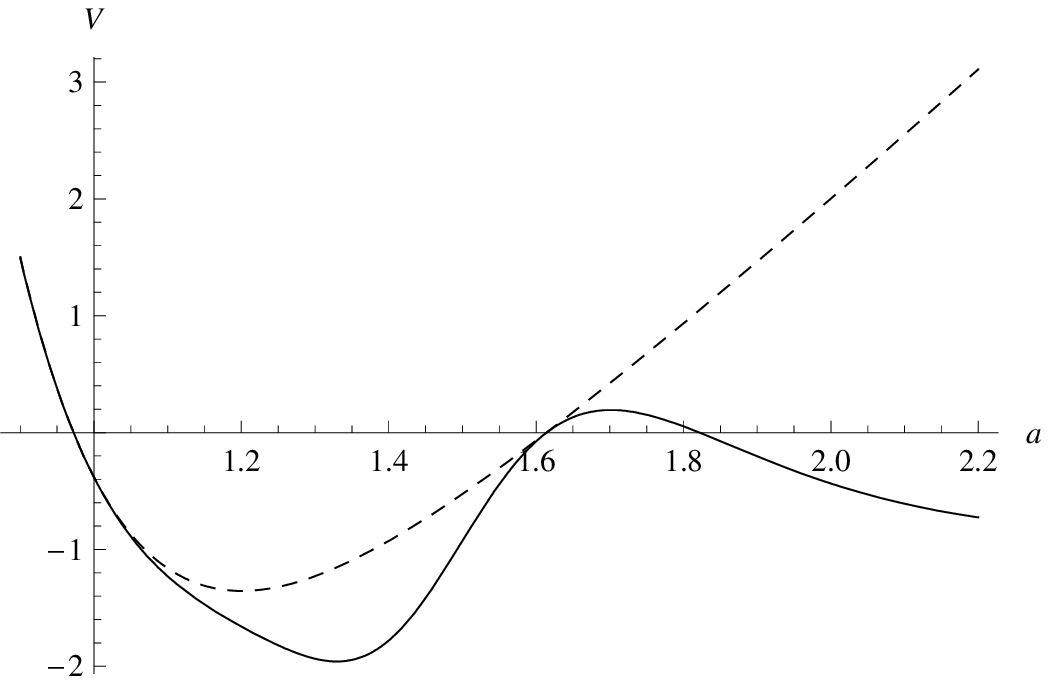}
   \includegraphics[scale=.75]{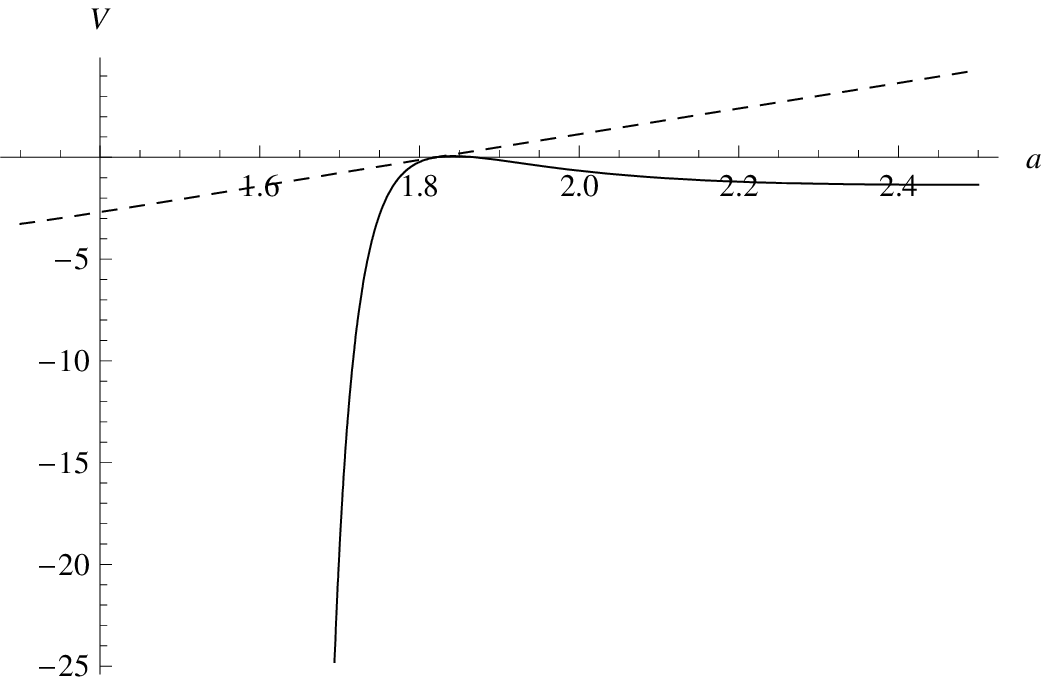}
\vspace{0cm}
        \caption{
For the same set of parameters as in the Fig. 1 and 2,
the plots on
the near-horizon behaviors of the metric function and potential
are shown.
On the top panel, the plot for the extremal case is shown.
In the remaining three plots,
from the top to the bottom panels,
the values of the black hole mass which we choose
become larger.
In each panel, the solid and dashed lines represent the behavior of
$\hat V$ and $\hat f$, respectively. }
   \end{center}
 \end{minipage}
\end{figure}
In each panel, the solid and dashed lines
represent the behavior of $\hat V$ and $\hat f$, respectively.
The top panel corresponds to the extremal case, and
the remaining three panels to the case of two-horizon
(the values of the mass chosen become larger,
from the top to the bottom).
One can see that in any case
a bouncing can happen just in front of the outer horizon,
although the height of the potential barrier is being smaller,
leading to longer bouncing time for larger values of mass.

As the reference, in Fig. 4,
we show the behavior of potential in neglecting
the electrostatic energy (namely setting $\hat G=0$ by hand).
These results have been essentially shown in Refs. \cite{rn1,rn2,rn3,bibh}.
\begin{figure}
\begin{minipage}[t]{.45\textwidth}
   \begin{center}
    \includegraphics[scale=.75]{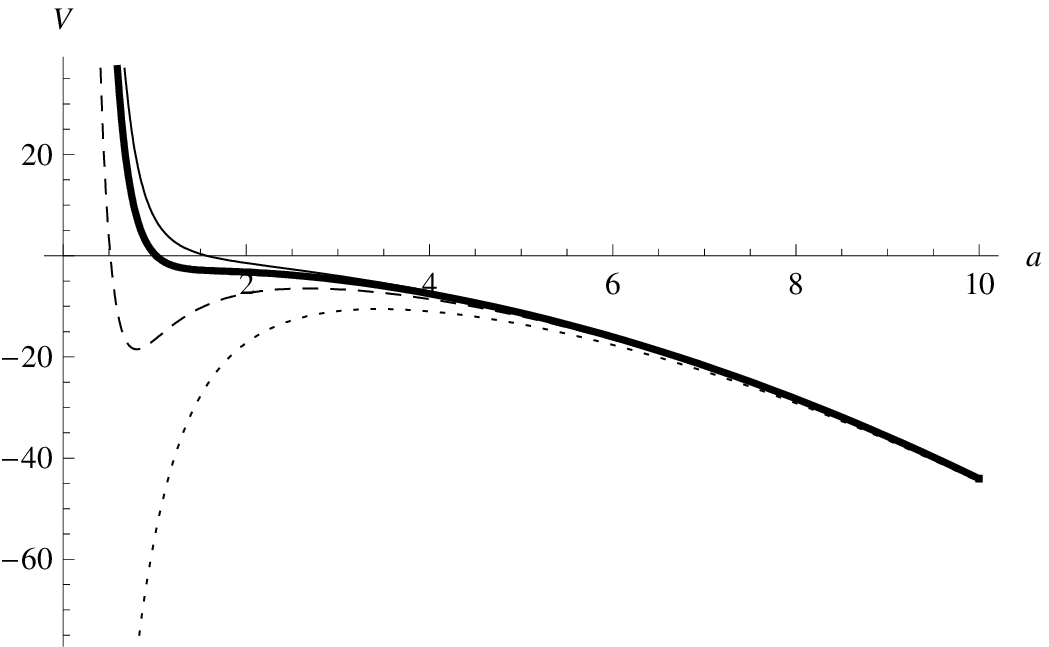}
\vspace{0cm}
        \caption{
Typical behaviors of the potential $\hat V$
are shown as the function of $\hat a$, by setting $\hat G=0$ by hand. The solid, thick, dashed and dotted curves show no horizon ($m^2<m_{\rm ext}^2$),
extremal ($m^2=m_{\rm ext}^2$), two-horizon ($A>m^2>m_{\rm ext}^2$)
and single horizon ($A<m^2$) cases. We have chosen $n=4$, $\hat \beta=10$,
$\hat q=3$ and $x=1.1$ (RS-tuning).
}
   \end{center}
 \end{minipage}
\end{figure}
Except for the final single-horizon case ($m^2>A$),
the domain wall could undergo a bounce inside the (outer) horizon.
Our results showed that the inclusion of the electrostatic energy
brings the bouncing point outside the horizon.

The feature of the potential is unchanged for difference choices
of $x$ as long as $x\geq 1$.
Also for $K=\pm 1$,
the results reaming essentially the same.
The bouncing of the domain wall universe
always happens just outside the outer horizon
and height of the barrier becomes smaller for larger values of
the black hole mass,
leading to longer time scales for bouncing.


\vspace{0.2cm}

\paragraph*{Summary:}
In this Letter,
we discussed the dynamics of
a domain wall universe in the
Einstein-Born-Infeld (EBI) theory.
We assumed that the spacetime is
asymptotically anti-de Sitter (AdS).
In the previous works,
the electrostatic energy
of the wall is not taken into account.
In this work, we have taken it into consideration,
which is determined through the $U(1)$ junction condition.

We obtained the effective Friedmann equation on the wall
in the EBI theory.
There are four possible spacetime structures, i.e.,
those with no horizon (naked singularity),
the extremal one,
those with two horizons (as the Reissner-Nordstr$\rm{\ddot o}$m black hole),
and those with a single horizon (as the Schwarzshild black hole) cases.
We find that a cosmological bounce always can happen
{\it outside} the (outer) horizon.
The height of the barrier between the bouncing point
and horizon in the effective potential
is being smaller for larger values of the black hole mass,
which leads to the longer bouncing time.
These results are in contrast to the results obtained in previous works,
suggesting that
regular bounce can happen {\it inside} the horizon except for $m^2>A$.

In all the cases discussed in this Letter,
at the later times, the contribution of the electrostatic energy
falls off very rapidly proportional to $a^{-7}$ and
the domain wall cosmology reduces to the one in the neutral
black hole background, which include the case of the Randall-Sundrum model.

Finally, we shall mention the stability of the system discussed 
in this Letter.
We speculate that our system is stable 
for most possible choices of parameters,
in terms of the stability of 
asymptotically AdS, charged and static black hole solutions.
To our knowledge, there has been no work
which investigated the dynamical stability of the charged,
asymptotically AdS black hole solutions in the EBI theory.
In the case of the $D\geq 5$-dimensional Einstein-Maxwell theory,
in Ref. \cite{ki}
the stability of Reissner-Nordstr$\rm{\ddot o}$m AdS black hole
in $D\geq 5$ against the vector and tensor perturbations
was shown.
The stability against the scalar perturbations
in $D=5,6,\cdots,11$-dimensional EM theory was shown
for all the parameters of charge and cosmological constant 
in Ref. \cite{sta}.
In addition, in terms of the thermodynamical arguments, 
in Ref. \cite{bibh2},
it was shown that asymptotically AdS black holes in EBI theory
are thermodynamical stable, always for $K=0,-1$,
and for $K=+1$ if the BI parameter $\beta$
is larger than some critical value $\beta_c$.
These facts suggest that the asymptotic AdS black hole solutions in 
the EBI theory are also dynamically stable 
for at least most possible choices of parameters.
Thus, the system discussed in this Letter is also expected 
to be stable.

\section*{Acknowledgements}
This work was supported by the Korea Science and Engineering Foundation(KOSEF)
grant funded by the Korea government (MEST) through the Center for Quantum
Spacetime (CQUeST).
W.L. was supported by the Korea Research Foundation Grant funded by the Korean Government (MOEHRD)(KRF-2007-355-C00014).

\appendix

\end{document}